\documentclass{sig-alternate-05-2015}

\usepackage[
  pass,
]{geometry}

\usepackage{etoolbox}
\makeatletter
\patchcmd{\maketitle}{\@copyrightspace}{}{}{}
\makeatother

\usepackage[utf8]{inputenc}
\usepackage[english]{babel}

\usepackage{hyperref}

\hypersetup{
    colorlinks=true,
    citecolor=magenta,
    linkcolor=blue,
    filecolor=magenta,      
    urlcolor=cyan
}
\usepackage{breakurl}
\usepackage{color}
\usepackage{array}
\usepackage[caption=false,font=footnotesize]{subfig}
\usepackage{multirow}
\usepackage{xspace}
\usepackage{float}
\usepackage{times}
\usepackage{setspace}
\usepackage{balance}
\usepackage{hyphenat}

\newcommand{\tchi}[0]{\emph{chi-15}\xspace}
\newcommand{\tsj}[0]{\emph{sj-12}\xspace}
\newcommand{\tmawi}[0]{\emph{mawi-15}\xspace}
\newcommand{\tfb}[0]{\emph{fb-web}\xspace}

\title{Relaxing state-access constraints in stateful programmable data planes}

\author{\rm{
Carmelo Cascone$^{\ddagger}$,
Roberto Bifulco$^{\ast}$,
Salvatore Pontarelli$^{+}$},
Antonio Capone$^{\ddagger}$
\\ 
\rm{
$^{\ddagger}$ Politecnico di Milano,
$^{\ast}$ NEC Laboratories Europe,
$^{+}$ CNIT/Univ. Roma Tor Vergata}
\\
\rm{
carmelo.cascone@polimi.it, roberto.bifulco@neclab.eu,
salvatore.pontarelli@uniroma2.it, antonio.capone@polimi.it}}

\begin{document}

\maketitle

\begin{abstract}
    Supporting the programming of stateful packet forwarding functions in hardware has recently attracted the interest of the research community. When designing such switching chips, the challenge is to guarantee the ability to program functions that can read and modify data plane's state, while keeping line rate performance and state consistency. Current state-of-the-art designs are based on a very conservative \emph{all-or-nothing} model: programmability is limited only to those functions that are guaranteed to sustain line rate, with any traffic workload. In effect, this limits the maximum time to execute state update operations.

    In this paper, we explore possible options to relax these constraints by using simulations on real traffic traces. We then propose a model in which functions can be executed in a larger but bounded time, while preventing data hazards with memory locking. We present results showing that such flexibility can be supported with little or no throughput degradation.
\end{abstract}


\section{Introduction}
Processing large network traffic loads requires the realization of complex algorithms in the network data plane, both to implement smart traffic forwarding policies at line rate or to offload processing that used to happen on general purpose CPUs. To meet the required performance levels while still providing the ability to quickly adapt the algorithms to new emerging needs, a new generation of programmable network switches and interfaces has been proposed~\cite{rmt, flexnic}.

In contrast with previous work on active networks \cite{activenet}, network processors \cite{intelixp} and software network functions~\cite{clickos}, this new generation of programmable data planes can provide programmability without paying the cost of a lower forwarding performance. RMT~\cite{rmt} is an example of a hardware design that can achieve high throughput while providing a programmable stateless Match-Action Table (MAT) abstraction, similar to the one provided by OpenFlow~\cite{openflow}, but with programmer-defined protocol fields and forwarding actions.
Recently, Sivaraman et al.~\cite{domino} demonstrated that it is also possible to implement a programmable high performance stateful data plane in hardware, provided that strict constraints on the per-packet execution time are met. Here, the challenge is to guarantee the ability to program stateful algorithms that read and modify data plane's state, while keeping line rate performance and state consistency.

More specifically, the implementation of a high performance hardware data plane requires processing packets in parallel. Typically, this is achieved with a pipeline design. Each pipeline's stage performs a few operations on a packet, and all the stages are executed in parallel. At each tick of the hardware's clock, packets are moved to the next pipeline's stage, the packet in the last stage exits the pipeline and a new packet enters in the first stage\footnote{While the data plane emits one packet per clock cycle, it is in fact processing a number of packets in parallel.}. The length of the pipeline finally defines the number of packets actually processed at the same time, in a given clock cycle. 

When state read and write operations are quick enough to be executed in a single pipeline's stage, i.e., in a clock cycle, the state consistency problem is inherently solved, while a \emph{data hazard} arises when more complex computations are required.
In fact, complex computations require longer time to be executed, therefore, they may not be completed within a single clock cycle and are instead split to be executed over multiple pipeline's stages. Since state is typically read in the first stage and written back, after modification, in the last stage, there is a risk the first stage may read an inconsistent state when a new packet enters the pipeline. That is, the read state is going to be invalidated by a result written back in a later stage.

In \cite{domino}, line rate performance are guaranteed by ensuring that state read and write operations happen within the same clock cycle, and that no state is shared between pipeline's stages. The cost of this design is the inability to express complex operations that take longer to complete. 

In this paper, we explore options to relax this constraint by making two observations. First, data planes are usually dimensioned for the worst case scenario, that is, processing minimum size packets at full line rate capacity. However, a data plane pipeline performs algorithms only acting on packets' header. For a given line rate, larger packet sizes actually mean a lower rate of packet headers to process. Hence, more time per packet header can be used to execute the pipeline operations.
Second, to process a given packet, many algorithms access only a subset of the overall data plane's state. If the pipeline stages access different portions of the state, then the risk of a data hazard is limited to the risk of having in the pipeline two or more packets whose processing requires access to the same portion of the state.

To check the actual data hazard probability when taking into account our observations, we designed a trace-based simulator and run it using real traffic traces, from both carrier and data center networks. 
To model state accesses, we observe that a common practice for many data plane algorithms is to deal with a network flow-level abstraction. Thus, we assume that only the processing of packets belonging to the same flow requires access to a common portion of the state. 
Our findings confirm that, in most cases, there is just a little probability of incurring in a data hazard even if state read and write operations happen in different clock cycles. Of course, such probability depends on the packet size and network flows distributions, as well as on the aggregation level of the network flow definition.  

Given our findings, we provide as second contribution a sketch of a pipeline design that avoids such data hazards by stalling the pipeline, when needed. 
Our simulations show that such a design is able to provide line rate throughput, despite the stalls, even extending the time between state read and write to several clock cycles. Furthermore, our design introduces little overhead in terms of memory and circuitry complexity when compared to state-of-the-art solutions. On the flip side of the coin, we are unable to provide line rate throughput in all the cases, therefore the data plane performance is dependent on the actual traffic load properties. Moreover, stalling the pipeline requires the introduction of small queues at the entrance of the pipeline stages. Dimensioning such queues introduces a new variable in the design space: small queues may provide lesser throughput, while big ones may introduce significant latency to the packet forwarding.

\section{Background}
\label{sec: background}
This section presents the data plane architecture we use as our reference model throughout the paper.

\begin{figure}
    \centering
    \includegraphics[clip,width=\columnwidth]{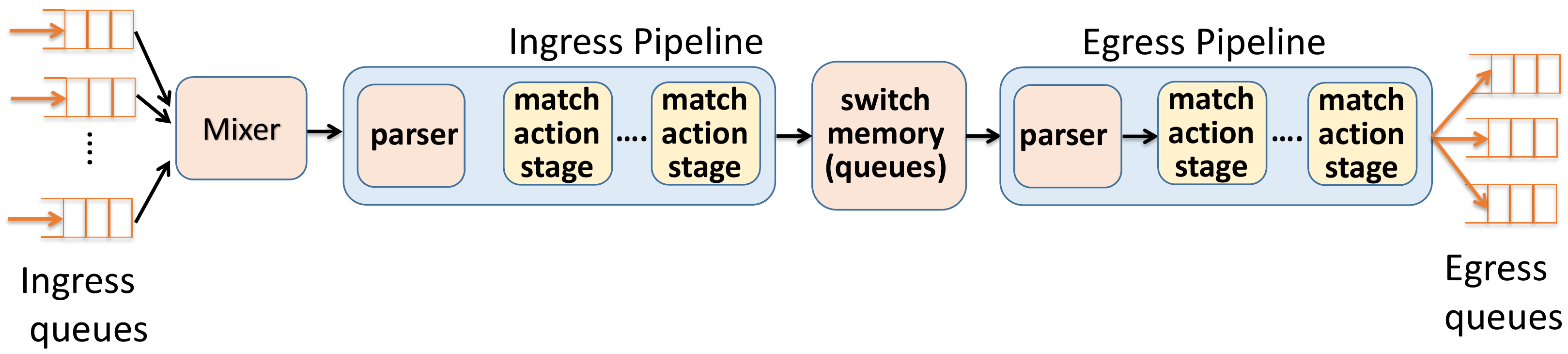}
    \caption{Programmable data plane architecture} 
    \label{fig:arch1}
\end{figure}

Data plane solutions such as RMT \cite{rmt}, Intel’s FlexPipe \cite{flexpipe}, and Cavium’s XPliant Packet Architecture \cite{cavium} implement a high level architecture similar to the one sketched in Figure~\ref{fig:arch1}.
In such architectures, packets received from the input ports are stored (enqueued) in a per-port queue and served, with a round robin policy, by a mixer that feeds the packets to an ingress pipeline.
After the ingress pipeline, the packets are stored in a common data memory. A scheduler selects which packets should be transmitted to the egress queues. A packet selected for forwarding is first processed by an egress pipeline, which is in principle similar to the ingress one, before being finally transmitted to the egress queues\footnote{While in this paper we focus on a store-and-forward mode of operation, our considerations are applicable also to data planes that work in a cut-through configuration.}.

Ingress and egress pipelines are composed by a programmable packet parser~\cite{rmt} and by a variable number of Match-Action Table (MAT) elements. For each new packet, the parser extracts the \emph{headers} that are then processed by the MATs elements. 
A MAT element implements itself a pipeline, whose architecture may sensibly change from one implementation to the other~\cite{hpsr, domino}. In this paper, we focus on the MAT element's pipeline, since it is in here that stateful operations are executed.

For the sake of our discussion, we can ignore most of the details of the actual MAT element and model its internal pipeline using just a sequence of stages plus memory. Each stage performs a limited amount of operations, such as read and write from/to memory, or some sort of computation. 
Since a complex operation can be split in a number of simpler operations executed in multiple stages, the number of stages finally defines the complexity of the operations that can be implemented by the MAT element.
Also, since each stage adds a clock cycle to the latency, the number of stages directly impacts the forwarding latency of a packet traversing the MAT element. Finally, and most importantly for the state consistency problem, the number of stages between a memory read and a memory write operations has important implications on the probability of incurring in a race condition. Intuitively, the longer is the time to process a value read from memory, before writing it back, the more probable is the reception of a new packet whose processing requires access to that same memory area. 
Without loss of generality and to simplify our exposition, in this paper we assume that the first MAT element's  pipeline's stage reads from memory, while the last one writes back to it.

\subsection{Rethinking design assumptions}
Recently proposed programmable stateful data planes, such as Banzai~\cite{domino}, eliminate data hazards by executing memory read and write operations within the same stage, i.e., in one clock cycle. Such a design derives from the worst case assumption that all packets have minimum size, that they arrive back-to-back, i.e., with no inter-packet gaps, and that they all need to access the same memory area. 
More specifically, let's consider a data plane with a throughput of 640 Gb/s, with a chip clocked at 1 GHz, as it is the case of RMT~\cite{rmt}. 
We can assume that packets are read in chunk of at most 80 bytes (i.e., 80 $\times$ 8 bit $\times$ 1 GHz $=$ 640 Gb/s) when entering the data plane's pipeline. Consequently, it will take 1 clock cycle to read packets with minimum size $\le 80$ bytes, while it will take more cycles to read longer packets, e.g., 19 for 1500 bytes.  

As mentioned earlier, the data plane's pipeline is dimensioned to accept a new packet's headers at each clock cycle. However, even when all packets arrive back-to-back, the variability of the packet size will cause the pipeline to experience one or more idle cycles. For example, in the case of maximum size packets of 1500 bytes, the pipeline will receive a new packet's headers at intervals of 19 clock cycles\footnote{While in principle it is possible to widen the pipeline data-path to reduce the maximum number of clock cycles to process a packet, and to increase the throughput, the routing of such a big number of parallel wires prevents several technological challenges that actually limit the maximum data-path width. For example in RMT\cite{rmt} it is explicitly mentioned that the  data-path is limited due to these technology constraints. Furthermore, the maximum achievable throughput is finally limited by the network interfaces speed.}.

We observe that packets produced by today's application have very variable size distributions. E.g., spanning from 64 bytes to 1500 bytes, in a typical case. Which could leave some space for relaxing the constraint on the memory read and write operations, when dealing with non corner-case traffic loads.

\subsection{Per-flow concurrency}

The second observation is that data plane state can be categorized in two types: \emph{global state} and \emph{flow state}. The first type is state that is shared among all packets, with no distinction, while flow state is shared only by packets of the same flow. We do not put any restriction on the definition of a flow, for example it could be the TCP or UDP 5-tuple, the IP address source-destination pair, the destination IP address, a portion of the latter, e.g. the first 16 bits, or a switch-dependent metadata, such as the packet's ingress or egress port.

Usually packet processing functions use a combination of the two, or only one. For example, a stateful firewall needs to maintain state for each TCP connection. A source NAT (SNAT) that dynamically translates the source IP address and port of outgoing connections, needs to maintain both flow state and global state: flow state for each L4 connection in order to distinguish between packets of new or existing connections, and in the case of new ones, it needs to pick a source address and port from a pool of available ones. Maintaining a pool of addresses and ports, e.g. using a stack, is an example of global state. Advanced load balancing schemes such as CONGA \cite{conga} maintain state at different aggregation levels: i) the 5-tuple in order to distinguish between flowlets, i.e. burst of packets, of the same L4 connection, ii) tunnel ID to maintain real-time utilization levels of several paths and iii) global state to maintain the best path among all available ones, to assign new flowlets to.

The notion of flow state is at the base of existing abstractions such as OpenState \cite{openstate} and FAST \cite{fast} which both extends the processing of a MAT with stateful capabilities. In both approaches, before the packet is processed by a MAT, a flow key is derived by looking solely at a subset of header fields decided by a programmer at configuration-time. Regardless of the specific operation to compute the flow key, in both approaches the operation on state that a programmer can define are limited to the portion of the memory associated to the packet's flow key. Indeed, if we imagine the state as an array, where each cell contains the state of a specific flow, it follows that multiple packets accessing and modifying different cells can be processed in parallel, with no harm for consistency. If the flow key is used as an unique index to access the state array, then packets with different flow keys can be processed in parallel.

\begin{table}
\centering
\caption{Packet traces used in simulations}
\vspace{2mm}
\footnotesize
\resizebox{\columnwidth}{!}{
\begin{tabular}{llllll}
\hline
\multirow{2}{*}{\textbf{Trace}} & \multirow{2}{*}{\textbf{Source}}           & \multirow{2}{*}{\textbf{Num pkts}} & \multicolumn{3}{l}{\textbf{Num flows per 1m pkts}}          \\
                                &                                            &                                    & \textit{5-tuple} & \textit{ipdst} & \textit{ipdst/16} \\ \hline
chi-15                          & CAIDA \cite{chi15}                       & 3.5b                               & 100.6k             & 57.7k            & 4.6k                \\ \hline
sj-12                           & CAIDA \cite{sj12}                        & 3.6b                               & 429k               & 17k              & 2k                  \\ \hline
mawi-15                         & MAWI \cite{mawilab}\cite{mawi15}       & 135m                               & 40.8k              & 17.3k            & 1.7k                \\ \hline
fb-web                          & Facebook \cite{fb-sigcomm}\cite{fbweb} & 447m                               & n/a                & n/a              & n/a                 \\ \hline
\end{tabular}
}
\label{table:trace summary}
\end{table}

\begin{figure}
\centering
\includegraphics[width=\columnwidth]{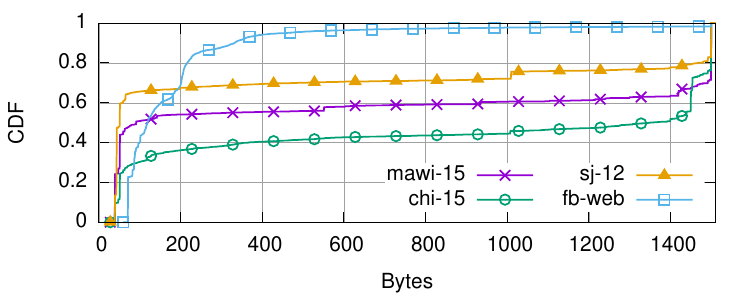}
\caption{Packet size cumulative distribution.} 
\label{fig:pkt size cdf}
\end{figure}

\section{Motivating experiments}
\label{sec: motivating experiments}

We start by evaluating the actual probability of generating data hazards when processing packets with stateful functions spanning many clock cycles. To do so we implemented a simulator that we feed we real traffic traces. The code of the simulator is available at \cite{opp-sim}.

\subsection{Traffic traces}
\label{sec:pkt traces}

We used 4 publicly available traffic traces (Table~\ref{table:trace summary}). Three traces (\tchi, \tsj, \tmawi) are taken from backbone carrier networks and one from a datacenter (\tfb). Each trace presents different characteristics, in terms of packet size (Figure~\ref{fig:pkt size cdf}) and number of flows when using different aggregation keys (Table~\ref{table:trace summary}). CAIDA publishes 1-hour long traces 4 times per year. We select \tchi as one representative of usual conditions as the packet size and number of flows is close to the majority of the other traces published by CAIDA in the recent years \cite{caida-stats}. The packet size presents a bimodal distribution, 30\% of packets have minimum size below 80 bytes, wile 50\% have larger size close to 1500 bytes. On the other hand, \tsj and \tmawi represent an abnormal situation. In both traces there is a prevalence of smaller packets and, in the case of \tsj, also an unusual large number of 5-tuples w.r.t. the number of distinct IP destination address.

Finally, \tfb includes packets collected from a Facebook's datacenter's cluster that serves web requests. As such it presents a predominance of small packets (80\% have size $<$ 200 bytes). It must be noted that this trace is the result of uniform sampling with rate 1:30k. As a consequence of the sampling, we were not able to count the number of distinct flows. The reason is that the probability that two consecutive packets belong to different flows is higher than the other traces, not because of the traffic characteristics but because packets are indeed distant in time (a distance potentially greater than the average flow life). Hence, we use \tfb only to measure the effects of the variable packet size, not the flow distribution.

\subsection{First results: fraction of data hazards}
\label{sec:data hazard}

We simulate the case of a stateful processing block comprising $N$ sequential (pipelined) stages, where each stage is executed in 1 clock cycle, and where the first stage reads from the memory while the last writes back. We call pipeline depth the length in clock cycles of the processing block, hence $N$. A data hazard is the event in which the first stage of the action pipeline processes a header, while another one is currently traveling in the same pipeline. Clearly, when $N=1$, there is no risk of data hazards.

We simulate the case when all packets are received at the switch back-to-back, hence line rate utilization is 100\% with no inter-packet gaps. Packets are read in chunks of 80 bytes (as in RMT), hence taking 19 clock cycles to read 1 packet of 1500 bytes. For $N \leq 18$, there is no risk of data hazard. Conversely, with small packets the pipeline will experience shorter idle gaps. In the worst case, the headers of minimum size packets arriving back-to-back, will be also processed back-to-back in the pipeline, hence causing a data hazard for any $N>1$. When considering per-flow concurrency, if two headers belonging to \emph{distinct} flows are processed back-to-back, this does \emph{not} generate a data hazard.

In doing simulations we process traffic in batches of 100k packets. For each batch we compute the fraction of data hazards (FDH) over the total number of clock cycles needed to process 100k packets (which depends on the packet size). To reduce simulation time, for each trace we select batches at a rate of 1:100, in other words we evaluate one batch of 100k consecutive packets every 10m packets. The observation here is that traffic characteristics vary slowly in a period of 10m packets, hence multiple batches close in time will produce similar results. For each trace, we extract the 99th percentile from all FDH samples. As an example, if for a given trace the 99th percentile of the FDH is 0.3, it means that in the 99\% of batches evaluated, the FDH was below 30\%.

\begin{figure}
    \centering
    \subfloat[][]{
        \includegraphics[trim={0 2mm 0 0},clip,width=\columnwidth]{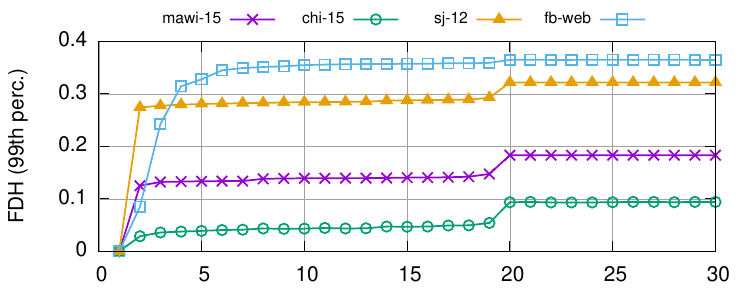}
        \label{fig:hazard 1F}
    }\\
    \subfloat[][]{
        \includegraphics[width=\columnwidth]{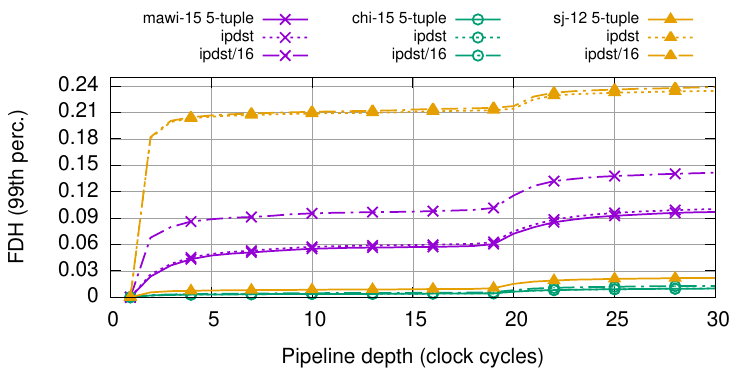}
        \label{fig:hazard MF}
    }
    \caption{Fraction of data hazards (FDH) w.r.t. increasing pipeline depth. In (a) all packets are considered belonging to the same flow, in (b) results are shown when aggregating packets per different flow keys.}
    \label{fig:my_label}
\end{figure}

Figure~\ref{fig:hazard 1F} shows the results for all traces when all packets are considered belonging to the same flow, i.e. accessing global state. Instead, in Figure~\ref{fig:hazard MF} FDH values are plotted for each trace when aggregating packets with different flow keys. As expected, the FDH greatly depends on the packet size distribution and flow keys, with smaller probability of hazards for traces with higher prevalence of larger packets, and for longer, i.e. finer, flow aggregation keys. such a \tchi. In the second case, per-flow concurrency affects the results. For example, with \tsj when considering state associated to distinct 5-tuples, the risk on incurring in a data hazard is way below the case when state is associated to distinct destination IP addresses. This result follows the flow distribution showed in Table~\ref{table:trace summary}. For all traces, using 5-tuples performs better than other flow keys. For \tchi, in all cases the FDH is around 1\%.

This result is important because it shows the probability of creating inconsistent state, and hence, if memory locking is a viable approach, the probability of incurring in such locking, thus affecting both throughput and latency.

\begin{figure}[t]
\centering
\includegraphics[width=1\columnwidth]{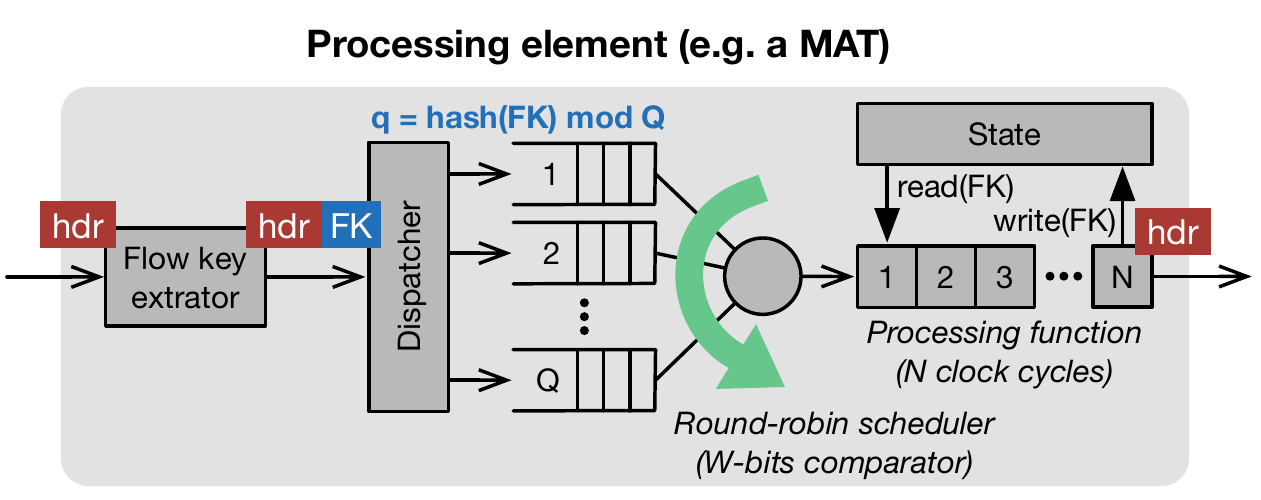}
\caption{Architecture of a stateful processing block with memory locking. Headers are queued based on their flow key (FK). A scheduler compares the FK of the head of each queue with what is traveling in the function pipeline, admitting only one header per flow.} 
    \label{fig:locking arch}
\end{figure}

\section{Approach: memory locking}
\label{sec:approach}

We propose here an approach to perform memory locking among packets competing to access the same memory portion. Locking is implemented by stalling the pipeline. That is, if two packets of the same flow arrive back-to-back, processing is paused for the second packet until the first one has left the pipeline. This already affects throughput by a factor of $1/N$ per flow, hence aggregate throughput is maximized when at least $N$ flows are active. Clearly, stalling calls for buffering which also introduces additional latency. We are interested in measuring the impact on throughput and latency when such locking approach is implemented.

We present a simple but effective pipeline design that implements stalling in order to prevent data hazards (Figure~\ref{fig:locking arch}). In our design, a stateful processing function spanning N clock cycles, is preceded by few $Q$ queues and a scheduler. For each packet's headers, a first block extracts a flow key (FK), then a dispatcher stores the headers in the $q$-th queue, where $q = hash(FK)\; mod\; Q$, thus preserving the processing order between packets of the same flow. Each queue can store at most $Q_{len}$ headers.

The scheduler decides which packet to admit in the processing pipeline by looking at the tip of each queue and comparing the head-of-line FK with the at most $N$ FKs currently traveling in the pipeline. The scheduler admits a header if its FK is \emph{not} currently in the pipeline. We assume the scheduler is work-conserving, meaning that all non-empty queues are compared at the same time, if at least one header can be served it will do so. To avoid starvation of a queue, the scheduler serves queues in a round-robin fashion, i.e. with cyclic priority.

We assume a FK can have arbitrary length of $FK_{len}$ bits, depending on the number of state memory cells available, for example $FK_{len} = 32$ bits for $2^{32}$ memory cells. As $Q < 2^{FK_{len}}$, multiple flows will end up sharing the same queue. Such an event may generate head-of-line blocking, in which all packets in a queue are held by the first one. Clearly, such a problem can be reduced by adding more queues, which has a cost in terms of silicon needed to implement both the queues and the scheduler.

For the scheduler to be work-conserving, it needs to compare all queues at the same time, hence increasing the number of wires with i) the number of queues and ii) the number of bits to compare for each queue. For this reason, to simplify the implementation of the scheduler's comparator, the FK is reduced to a smaller space of $W$ bits. This operation can be performed by the flow key extractor, which along the FK (needed later to access the state) extends the headers with a field $w$. In our experiments we compute $w = hash(FK)\; mod\; W$. For example with $W=4$, the scheduler is able to distinguish among $2^{4}=16$ flows. If $W < FK_{len}$ there will be different flows colliding onto the same value $w$, impacting performance. Flow collision also depends on the hash function, in all our experiments $hash() = crc16()$, which is a common feature in packet processing architectures. However, we do not investigate the impact of other hash functions on the distribution of FKs among the different queues and values of $w$.
 
\begin{table*}[!thb]
\centering
\caption{Clock cycle budget (and latency) when using memory locking}
{\footnotesize Maximum number of clock cycles (up to 30) per processing function, to sustain a given throughput. In all cases $W=4\; bits$. ``Global'' represents the case when packets need to access global state. Latency values are given for 1 Ghz clock frequency, i.e. 1 clock cycle = 1 ns.}\\
\vspace{1mm}
\label{table:cbudget}
\footnotesize
\resizebox{\textwidth}{!}{%
\begin{tabular}{ccc|cccc|cccc|cccc|c|}
\cline{4-16}
                                              &                                           &    & \multicolumn{4}{c|}{chi-15} & \multicolumn{4}{c|}{sj-12} & \multicolumn{4}{c|}{mawi-15} & fb-web \\ \hline
\multicolumn{1}{|c|}{Thrpt}                   & \multicolumn{1}{c|}{$Q_{len}$}            & $Q$& 5-tuple           & ipdst             & ipdst/16         & global       & 5-tuple           & ipst              & ipdst/16          & global         & 5-tuple           & ipdst             & ipdst/16       & global      & global        \\ \hline \hline
\multicolumn{1}{|c|}{\multirow{8}{*}{100\%}}  & \multicolumn{1}{c|}{\multirow{4}{*}{10}}  & 1  & 1                 & 1                 & 1                & 1            & 1                 & 1                 & 1                 & 1              & 1                 & 1                 & 1              & 1           & 1             \\ \cline{3-16} 
\multicolumn{1}{|c|}{}                        & \multicolumn{1}{c|}{}                     & 4  & 1                 & 1                 & 1                &              & 1                 & 1                 & 1                 &                & 1                 & 1                 & 1              &             &               \\ \cline{3-16} 
\multicolumn{1}{|c|}{}                        & \multicolumn{1}{c|}{}                     & 8  & 1                 & 1                 & 1                &              & 1                 & 1                 & 1                 &                & 1                 & 1                 & 1              &             &               \\ \cline{3-16} 
\multicolumn{1}{|c|}{}                        & \multicolumn{1}{c|}{}                     & 16 & 1                 & 1                 & 1                &              & 1                 & 1                 & 1                 &                & 1                 & 1                 & 1              &             &               \\ \cline{2-16} 
\multicolumn{1}{|c|}{}                        & \multicolumn{1}{c|}{\multirow{4}{*}{100}} & 1  & 20 (174ns)        & 20 (190ns)        & 21 (230ns)       & 8 (282ns)    & 4 (49ns)          & 1                 & 1                 & 1              & 2 (18ns)          & 2 (20ns)          & 2 (35ns)       & 1           & 2 (10ns)      \\ \cline{3-16} 
\multicolumn{1}{|c|}{}                        & \multicolumn{1}{c|}{}                     & 4  & 30 (175ns)        & 30 (192ns)        & 30 (320ns)       &              & 8 (152ns)         & 1                 & 1                 &                & 2 (12ns)          & 2 (14ns)          & 2 (25ns)       &             &               \\ \cline{3-16} 
\multicolumn{1}{|c|}{}                        & \multicolumn{1}{c|}{}                     & 8  & 30 (137ns)        & 30 (144ns)        & 30 (259ns)       &              & 8 (133ns)         & 1                 & 1                 &                & 2 (12ns)          & 2 (14ns)          & 2 (25ns)       &             &               \\ \cline{3-16} 
\multicolumn{1}{|c|}{}                        & \multicolumn{1}{c|}{}                     & 16 & 30 (122ns)        & 30 (126ns)        & 30 (221ns)       &              & 8 (123ns)         & 1                 & 1                 &                & 2 (11ns)          & 2 (14ns)          & 2 (24ns)       &             &               \\ \hline \hline
               
\multicolumn{1}{|c|}{\multirow{8}{*}{99.9\%}} & \multicolumn{1}{c|}{\multirow{4}{*}{10}}  & 1  & 8 (16ns)          & 8 (16ns)          & 8 (18ns)         & 4 (18ns)     & 2 (5ns)           & 1                 & 1                 & 1              & 1                 & 1                 & 1              & 1           & 2 (10ns)     \\ \cline{3-16} 
\multicolumn{1}{|c|}{}                        & \multicolumn{1}{c|}{}                     & 4  & 14 (33ns)         & 14 (31ns)         & 14 (38ns)        &              & 2 (4ns)           & 1                 & 1                 &                & 1                 & 1                 & 1              &             &               \\ \cline{3-16} 
\multicolumn{1}{|c|}{}                        & \multicolumn{1}{c|}{}                     & 8  & 16 (39ns)         & 15 (30ns)         & 16 (44ns)        &              & 2 (4ns)           & 1                 & 1                 &                & 1                 & 1                 & 1              &             &               \\ \cline{3-16} 
\multicolumn{1}{|c|}{}                        & \multicolumn{1}{c|}{}                     & 16 & 17 (37ns)         & 18 (43ns)         & 16 (42ns)        &              & 2 (5ns)           & 1                 & 1                 &                & 1                 & 1                 & 1              &             &               \\ \cline{2-16} 
\multicolumn{1}{|c|}{}                        & \multicolumn{1}{c|}{\multirow{4}{*}{100}} & 1  & 27 (568ns)        & 27 (618ns)        & 26 (605ns)       & 8 (282ns)    & 6 (143ns)         & 2 (86ns)          & 2 (84ns)          & 1              & 3 (42ns)          & 3 (48ns)          & 3 (100ns)      & 2 (60ns)    & 2 (10ns)     \\ \cline{3-16} 
\multicolumn{1}{|c|}{}                        & \multicolumn{1}{c|}{}                     & 4  & 30 (175ns)        & 30 (192ns)        & 30 (320ns)       &              & 15 (526ns)        & 2 (79ns)          & 2 (77ns)          &                & 4 (41ns)          & 4 (52ns)          & 4 (135ns)      &             &               \\ \cline{3-16} 
\multicolumn{1}{|c|}{}                        & \multicolumn{1}{c|}{}                     & 8  & 30 (137ns)        & 30 (144ns)        & 30 (259ns)       &              & 22 (731ns)        & 2 (79ns)          & 2 (78ns)          &                & 4 (38ns)          & 4 (50ns)          & 4 (131ns)      &             &               \\ \cline{3-16} 
\multicolumn{1}{|c|}{}                        & \multicolumn{1}{c|}{}                     & 16 & 30 (122ns)        & 30 (126ns)        & 30 (221ns)       &              & 25 (741ns)        & 2 (79ns)          & 2 (72ns)          &                & 4 (37ns)          & 4 (49ns)          & 4 (129ns)      &             &               \\ \hline \hline
              
\multicolumn{1}{|c|}{\multirow{8}{*}{99\%}}   & \multicolumn{1}{c|}{\multirow{4}{*}{10}}  & 1  & 21 (80ns)         & 21 (80ns)         & 21 (89ns)        & 7 (60ns)     & 3 (14ns)          & 1                 & 1                 & 1              & 2 (13ns)          & 2 (14ns)          & 1              & 1           & 2 (10ns)     \\ \cline{3-16} 
\multicolumn{1}{|c|}{}                        & \multicolumn{1}{c|}{}                     & 4  & 30 (142ns)        & 30 (148ns)        & 30 (184ns)       &              & 10 (45ns)         & 1                 & 1                 &                & 5 (34ns)          & 4 (31ns)          & 2 (18ns)       &             &               \\ \cline{3-16} 
\multicolumn{1}{|c|}{}                        & \multicolumn{1}{c|}{}                     & 8  & 30 (129ns)        & 30 (138ns)        & 30 (184ns)       &              & 11 (47ns)         & 1                 & 1                 &                & 6 (42ns)          & 4 (31ns)          & 2 (18ns)       &             &               \\ \cline{3-16} 
\multicolumn{1}{|c|}{}                        & \multicolumn{1}{c|}{}                     & 16 & 30 (116ns)        & 30 (122ns)        & 30 (180ns)       &              & 12 (47ns)         & 1                 & 1                 &                & 7 (52ns)          & 4 (31ns)          & 2 (18ns)       &             &               \\ \cline{2-16} 
\multicolumn{1}{|c|}{}                        & \multicolumn{1}{c|}{\multirow{4}{*}{100}} & 1  & 30 (950ns)        & 30 (922ns)        & 30 (1.1us)       & 9 (842ns)    & 8 (268ns)         & 2 (86ns)          & 2 (84ns)          & 2 (171ns)      & 9 (422ns)         & 8 (380ns)         & 5 (316ns)      & 4 (379ns)   & 2 (10ns)     \\ \cline{3-16} 
\multicolumn{1}{|c|}{}                        & \multicolumn{1}{c|}{}                     & 4  & 30 (175ns)        & 30 (192ns)        & 30 (320ns)       &              & 22 (1.1us)        & 3 (285ns)         & 3 (283ns)         &                & 24 (1.7us)        & 17 (1.3us)        & 7 (572ns)      &             &               \\ \cline{3-16} 
\multicolumn{1}{|c|}{}                        & \multicolumn{1}{c|}{}                     & 8  & 30 (137ns)        & 30 (144ns)        & 30 (259ns)       &              & 30 (1.9us)        & 3 (290ns)         & 3 (292ns)         &                & 30 (2.1us)        & 23 (2.2us)        & 8 (753ns)      &             &               \\ \cline{3-16} 
\multicolumn{1}{|c|}{}                        & \multicolumn{1}{c|}{}                     & 16 & 30 (122ns)        & 30 (126ns)        & 30 (221ns)       &              & 30 (940ns)        & 3 (296ns)         & 3 (293ns)         &                & 30 (1.3us)        & 25 (2.5us)        & 8 (759ns)      &             &               \\ \hline        

\end{tabular}
}
\end{table*}

\subsection{Silicon overhead}

We evaluate now the requirements in terms of silicon area of the added hardware blocks. There are two types of blocks to consider: i) the logic needed to realize the proposed locking scheme and ii) the logic blocks that will realize the stateful processing function.

The combinatorial logic complexity of the locking scheme is basically that of $Q \times N$ comparators each one of $W$ bits, that corresponds to few thousands of logic gates. An ASIC chip nowadays has more than $10^8$ logic gates, hence we consider this header negligible for small values of $Q$, $N$ and $W$. The memory requirements to implement buffering of headers is $H_{len} \times Q \times Q_{len}$ bits, where $H_{len}$ is the length in bits of the data path. With $H_{len} = 88$ bytes (80 for the header and 8 for the metadata), $Q=4$ and $Q_{len}=100$ it requires 35,2 KB of memory overhead for the queues, that is approximately 3.5\% of the memory overhead compared with the memory of a MAT stage in RMT \cite{rmt}.

For the second type of block, silicon overhead depends on the actual function implemented, as such we are not able to provide numbers. However, in current programmable ASIC switching technology 80\% of chip area is due to memory (TCAMs and the IO, buffer, and queue subsystem), and less than 20\% area is due to logic \cite{rmt}. As a result, we expect that supporting more complex processing functions will not be constrained by chip area. In other words, it seems it is easier to add more logic than to add memory.

\section{Trace-based results}
\label{sec:results}

We evaluated the proposed architecture using the same traffic traces presented in Section~\ref{sec:pkt traces}. When collecting performance metrics, we used the same approach described in Section~\ref{sec:data hazard}: traffic is processed in batches of 100k packets, with 10m packets distance between each batch; and 100\% line rate utilization. For each batch of packets we compute the \emph{throughput} as the fraction of packets served by the scheduler, over the total number of packets received; and the \emph{latency} as the number of clock cycles from when the packet is completely received to when it is served by the scheduler, i.e. it enters the function pipeline. For simplicity we consider that when $N=1$, i.e. no locking required, latency is 0. Latency is computed for each packet, for each batch we take the 99th percentile among all latency values, finally we take the maximum among all batches for a given trace. For example, a latency value of 5 means that in the worst case, 99\% of the packets experienced a latency of no more than 5 clock cycles, e.g. 5ns at 1Ghz.

We evaluated these metrics when varying the different parameters described in Section~\ref{sec:approach} for the different traces. We present here a subset of the results, a more detailed collection of results can be found at \cite{opp-sim}.

Table~\ref{table:cbudget} shows results in terms of clock cycle budget, which is the maximum number of clock cycles allowed for a stateful function to complete execution, while sustaining a given throughput. For example, to sustain 100\% throughput, using queues of size 10 (headers) does not provide any benefit, as the clock cycle budget is 1 for each trace and flow key. However, by adding more capacity to queues up to 100, budget improves even when using only 1 queue, allowing for functions spanning 20 clock cycles for all flow keys with \tchi, and 4 clock cycles with \tsj, but only when aggregating packets per 5-tuple. Clearly, long queues impact latency. Both clock cycle budget and latency improve if we can admit for a lower throughput of 99.9\%, i.e. allowing for 0.1\% drop probability. Clearly, reducing utilization (100\% in our experiments) reduces further the risk of drop while maintaining the same cycle budget.

\section{Discussion}
\label{sec:discussion}

\textbf{Issues with blocking architectures}
While the proposed solution enables the execution of more complex operations directly in the data plane, it implements a blocking architecture. That is, for particular workloads, the data plane is unable to offer line rate forwarding throughput. As a consequence, the processing programmed in the data plane should be adapted to the expected network load characteristics, for the line rate to be achievable. Our work helps in defining the boundaries of the achievable performance, for a given workload and set of operations.
An additional problem of the dependency on the workload is the possibility to exploit such dependency, e.g., to perform a denial of service attack on the data plane.
However, the ability to program stateful algorithms in the data plane should help in detecting and mitigating such exploitations at little cost.

\textbf{What can we do more with more clock cycles?}
We do not have yet a concrete example of application, however, if one can tolerate a blocking architecture, she can trade the complexity of investigating hardware circuit design for a specific function, e.g. to enforce atomic execution as in Banzai~\cite{domino}, with the possibility of using simpler but slower hardware blocks.
To the far end, we envision the possibility of using a general purpose packet processor, carefully programmed to complete execution in a longer, but bounded, clock cycle budget, with predictable performance when the traffic characteristics are known.
Finally, another option is that of supporting larger memories (e.g. DRAM) which have slower access times ($>$ 1 clock tick at 1GHz) to read and write values. The same multi-queue scheduling approach could be used to coordinate access to multiple parallel memory banks, where each bank is associated to a queue.

\section{Conclusion}
\label{sec:conclusion}

This paper presented a model for a packet processing pipeline which allows execution of complex functions that read and write data plane's state at line rate, when read and write operations are performed at different stages. Prevention of data hazards is performed by stalling the pipeline. By using simulations on real traffic traces from both carrier and datacenter networks, we show that such model can be applied with little or no throughput degradation. The exact clock cycle budget and latency depends on the packet size distribution (more with larger packets) and on the granularity of the flow key used to access state (more with longer flow keys, e.g. the 5-tuple). The code used for the simulations and additional results are available at \cite{opp-sim}.

\section*{Acknowledgements}

This work has been partly funded by the EU in the context of the H2020 ``BEBA'' project (Grant Agreement: 644122).

\balance

\bibliographystyle{abbrv}
\bibliography{biblio}

\begin{thebibliography}{10}

\bibitem{chi15}
The {CAIDA UCSD} anonymized internet traces - {Chicago} 2015-02-19.
\newblock \url{http://www.caida.org/data/passive/passive_2015_dataset.xml}.

\bibitem{sj12}
The {CAIDA UCSD} anonymized internet traces - {San Jose} 2012-11-15.
\newblock \url{http://www.caida.org/data/passive/passive_2012_dataset.xml}.

\bibitem{caida-stats}
The {CAIDA UCSD} statistical information for the {CAIDA} anonymized internet
  traces.
\newblock \url{http://www.caida.org/data/passive/passive_trace_statistics.xml}.

\bibitem{fbweb}
Facebook {FBFlow} dataset - cluster {B}.
\newblock \url{https://www.facebook.com/network-analytics}.

\bibitem{flexpipe}
{Intel FlexPipe}.
\newblock
  \url{http://www.intel.com/content/dam/www/public/us/en/documents/product-briefs/ethernet-switchfm6000-series-brief.pdf}.

\bibitem{intelixp}
{IXP4XX} product line of network processors.
\newblock
  \url{http://www.intel.com/content/www/us/en/intelligent-systems/previous-generation/intel-ixp4xx-intel-network-processor-product-line.html}.

\bibitem{mawi15}
{MAWILab} traffic trace - 2015-07-20.
\newblock
  \url{http://www.fukuda-lab.org/mawilab/v1.1/2015/07/20/20150720.html}.

\bibitem{opp-sim}
{OPP-SIM} source code and results repository.
\newblock \url{https://github.com/ccascone/opp-sim}.

\bibitem{cavium}
{XPliant} ethernet switch product family.
\newblock
  \url{http://www.cavium.com/XPliant-Ethernet-Switch-ProductFamily.html}.

\bibitem{conga}
M.~Alizadeh, T.~Edsall, S.~Dharmapurikar, R.~Vaidyanathan, K.~Chu,
  A.~Fingerhut, V.~T. Lam, F.~Matus, R.~Pan, N.~Yadav, and G.~Varghese.
\newblock Conga: Distributed congestion-aware load balancing for datacenters.
\newblock In {\em ACM SIGCOMM '14}.

\bibitem{openstate}
G.~Bianchi, M.~Bonola, A.~Capone, and C.~Cascone.
\newblock Openstate: Programming platform-independent stateful openflow
  applications inside the switch.
\newblock {\em ACM SIGCOMM CCR}, April 2014.

\bibitem{rmt}
P.~Bosshart, G.~Gibb, H.-S. Kim, G.~Varghese, N.~McKeown, M.~Izzard, F.~Mujica,
  and M.~Horowitz.
\newblock Forwarding metamorphosis: Fast programmable match-action processing
  in hardware for sdn.
\newblock In {\em ACM SIGCOMM '13}.

\bibitem{mawilab}
R.~Fontugne, P.~Borgnat, P.~Abry, and K.~Fukuda.
\newblock {MAWILab}: Combining diverse anomaly detectors for automated anomaly
  labeling and performance benchmarking.
\newblock In {\em ACM CoNEXT '10}.

\bibitem{flexnic}
A.~Kaufmann, S.~Peter, T.~Anderson, and A.~Krishnamurthy.
\newblock {FlexNIC}: rethinking network {DMA}.
\newblock In {\em USENIX HotOS '15}.

\bibitem{clickos}
J.~Martins, M.~Ahmed, C.~Raiciu, V.~Olteanu, M.~Honda, R.~Bifulco, and
  F.~Huici.
\newblock Clickos and the art of network function virtualization.
\newblock In {\em USENIX NSDI'14}.

\bibitem{openflow}
N.~McKeown, T.~Anderson, H.~Balakrishnan, G.~Parulkar, L.~Peterson, J.~Rexford,
  S.~Shenker, and J.~Turner.
\newblock Openflow: Enabling innovation in campus networks.
\newblock {\em ACM SIGCOMM CCR}, April 2008.

\bibitem{fast}
M.~Moshref, A.~Bhargava, A.~Gupta, M.~Yu, and R.~Govindan.
\newblock Flow-level state transition as a new switch primitive for sdn.
\newblock In {\em ACM HotSDN '14}.

\bibitem{hpsr}
S.~Pontarelli, M.~Bonola, G.~Bianchi, A.~Capone, and C.~Cascone.
\newblock Stateful {Openflow}: Hardware proof of concept.
\newblock In {\em IEEE HPSR '15}.

\bibitem{fb-sigcomm}
A.~Roy, H.~Zeng, J.~Bagga, G.~Porter, and A.~C. Snoeren.
\newblock Inside the social network's (datacenter) network.
\newblock In {\em ACM SIGCOMM '15}.

\bibitem{domino}
A.~Sivaraman, A.~Cheung, M.~Budiu, C.~Kim, M.~Alizadeh, H.~Balakrishnan,
  G.~Varghese, N.~McKeown, and S.~Licking.
\newblock Packet transactions: High-level programming for line-rate switches.
\newblock In {\em ACM SIGCOMM '16}.

\bibitem{activenet}
D.~L. Tennenhouse and D.~J. Wetherall.
\newblock Towards an active network architecture.
\newblock In {\em DARPA Active Networks Conference and Exposition}, 2002.

\end{thebibliography}

\end{document}